\documentclass[10pt]{article}
\usepackage{caption}
\usepackage{graphicx}
\usepackage{subcaption}
\usepackage[inline]{enumitem}
\usepackage{authblk}
\usepackage{xcolor}

\usepackage{parskip}

\usepackage[binary-units=true]{siunitx}
\AtBeginDocument{\DeclareSIUnit{\Wh}{Wh}}

\usepackage{glossaries}
\newacronym{pse}{PSE}{Power Sourcing Equipment}
\newacronym{pd}{PD}{Powered Device}
\newacronym{usrp}{USRP}{Universal Software Radio Peripheral}
\newacronym{rpi}{RPi}{Raspberry Pi}
\newacronym{poe}{PoE}{Power over Ethernet}
\newacronym{ptp}{PTP}{Precision Time Protocol}
\newacronym{vlc}{VLC}{Visible Light Communication}
\newacronym{vlp}{VLP}{Visible Light Positioning}
\newacronym{tsn}{TSN}{Time-Sensitive Networking}
\newacronym{sdr}{SDR}{Software-Defined Radio}
\newacronym{los}{LoS}{Line-of-Sight}
\newacronym{pps}{PPS}{Pulse Per Second}
\newacronym{gpclk}{GPCLK}{General Purpose CLock}
\newacronym{ros}{ROS}{Robot Operating System}
\newacronym{lco}{LCO}{Lithium Cobalt Oxide}
\newacronym{tof}{ToF}{Time-of-Flight}
\newacronym{ntp}{NTP}{Network Time Protocol}
\newacronym{soc}{SoC}{State of Charge}
\newacronym{uhd}{UHD}{USRP Hardware Driver}
\newacronym{pcb}{PCB}{Printed Circuit Board}
\newacronym{pll}{PLL}{phase-locked loop}
\newacronym{os}{OS}{Operating System}
\newacronym{hat}{HAT}{Hardware Attached on Top}
\newacronym{gpu}{GPU}{Graphical Processing Unit}
\newacronym{cpu}{CPU}{Central Processing Unit}
\newacronym{if}{IF}{intermediate frequency}
\newacronym{fpga}{FPGA}{Field-Programmable Gate Array}
\newacronym{rfic}{RFIC}{radio-frequency integrated circuit}
\newacronym{ssd}{SSD}{solid state drive}

\usepackage{placeins}

\usepackage[backend=biber,style=ieee]{biblatex}
\AtBeginBibliography{\footnotesize}

\usepackage{url}

\usepackage[super]{nth}

\makeatletter
\def\blfootnote{\gdef\@thefnmark{}\@footnotetext}
\makeatother

\providecommand{\keywords}[1]
{
  \small	
  \textbf{\textit{Keywords---}} #1
}

\begin{document}

\baselineskip 12pt

\begin{center}
{\Large \textbf{A Primer on Techtile: An R\&D Testbed for Distributed Communication, Sensing and Positioning}\par}
\vspace{5mm}
\begin{tabular}{cccc}  Gilles Callebaut  & Jarne Van Mulders & Geoffrey Ottoy &  Liesbet Van der Perre \\[1.2\baselineskip]
\multicolumn{4}{c}{ESAT-DRAMCO, Ghent Technology Campus, KU Leuven} \\
\multicolumn{4}{c}{9000 Ghent, Belgium} \\[1.2\baselineskip]
\multicolumn{4}{c}{%
 \texttt{gilles.callebaut@kuleuven.be}}
\end{tabular} 
\end{center}
\vspace{5mm}


\begin{abstract}
\noindent The Techtile measurement infrastructure is a multi-functional, versatile testbed for new communication and sensing technologies relying on fine-grained distributed resources. The facility enables experimental research on hyper-connected interactive environments and validation of new wireless connectivity, sensing and positioning solutions. It consists of a data acquisition and processing equipment backbone and a fabric of dispersed edge computing devices, \glspl{sdr}, sensors, and LED sources. These bring intelligence close to the applications and can also collectively function as a massive, distributed resource. Furthermore, the infrastructure 
allows exploring more degrees and new types of diversity, i.e., scaling up the number of elements, introducing `3D directional diversity' by deploying the distributed elements with different orientations, and `interface diversity' by exploiting multiple technologies and hybrid signals (RF, acoustic, and visible light).
\end{abstract}
\keywords{Testbed, 6G, Software-Defined Radio, Precision Time Protocol, Power-over-Ethernet, RadioWeaves}

\section{Techtile -- A Teaser}
\begin{figure}[h!]
\centering
    \centering
    \includegraphics[width=0.6\linewidth,trim={3cm 10.5cm 13cm 2cm},clip]{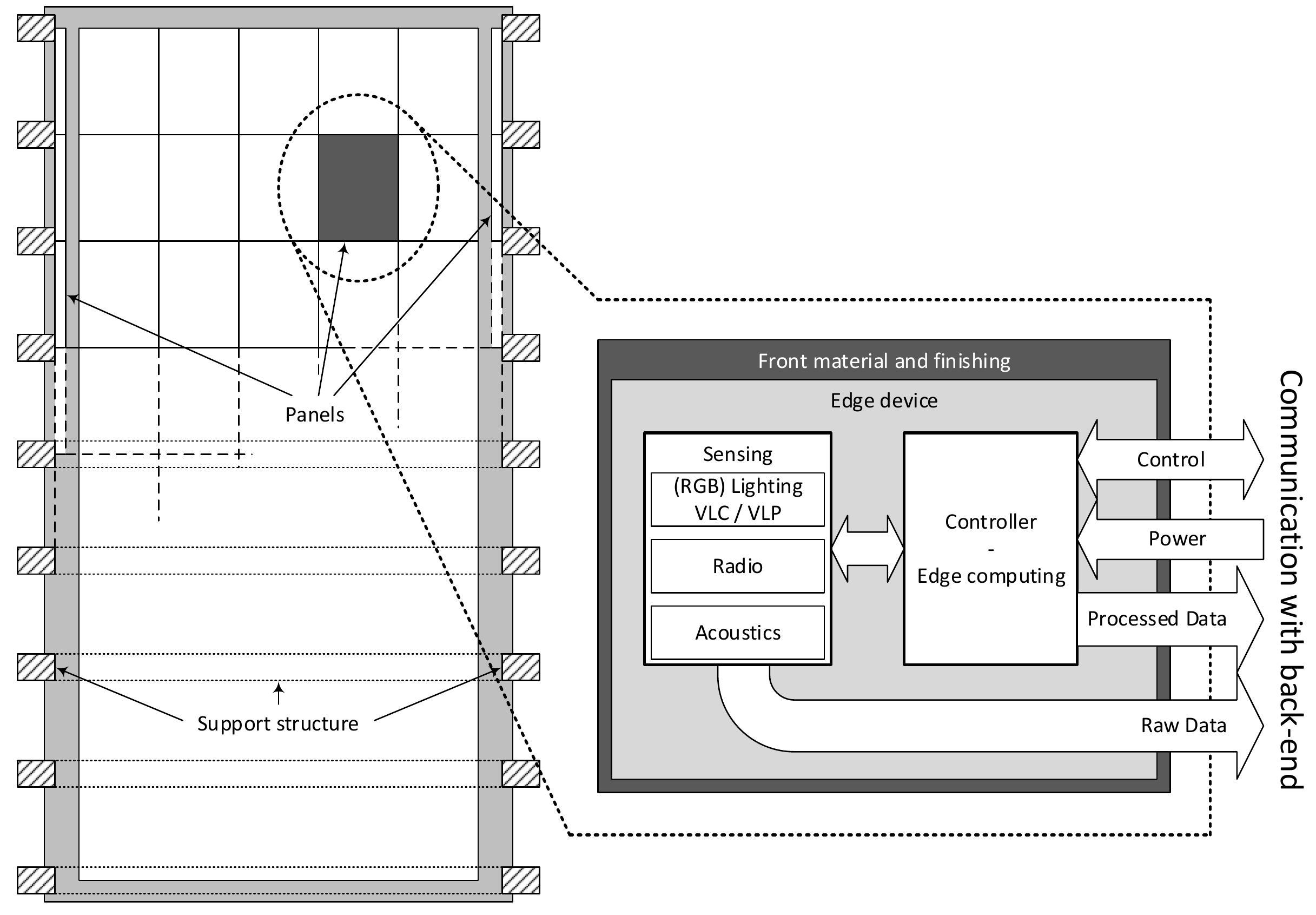}%
    \caption{\small Left: Support structure (top view) -- Right: Tile with embedded edge device. The support structure is designed to support 140 detachable panels.}%
    \label{fig:System}
\end{figure}

As envisioned in~\cite{VanderPerreLiesbet2019Rfec}, new wireless access infrastructures will be integrated in existing structures, bringing them in closer proximity of devices and eventually becoming truly ubiquitous. The Techtile infrastructure (Figure~\ref{fig:System}) is the first testbed capable of evaluating this RadioWeaves concept.
To support this, the measurement room or --more general-- the testbed, hosts 140 detachable tiles of equal size on the walls, floor, and ceiling. A versatile set of equipment can be mounted on these panels, effectively embedding the electronics into the room.


\begin{figure}[!htb]
      \centering
      \begin{subfigure}[t]{0.6\textwidth}
         \centering
         \includegraphics[height=2.2in]{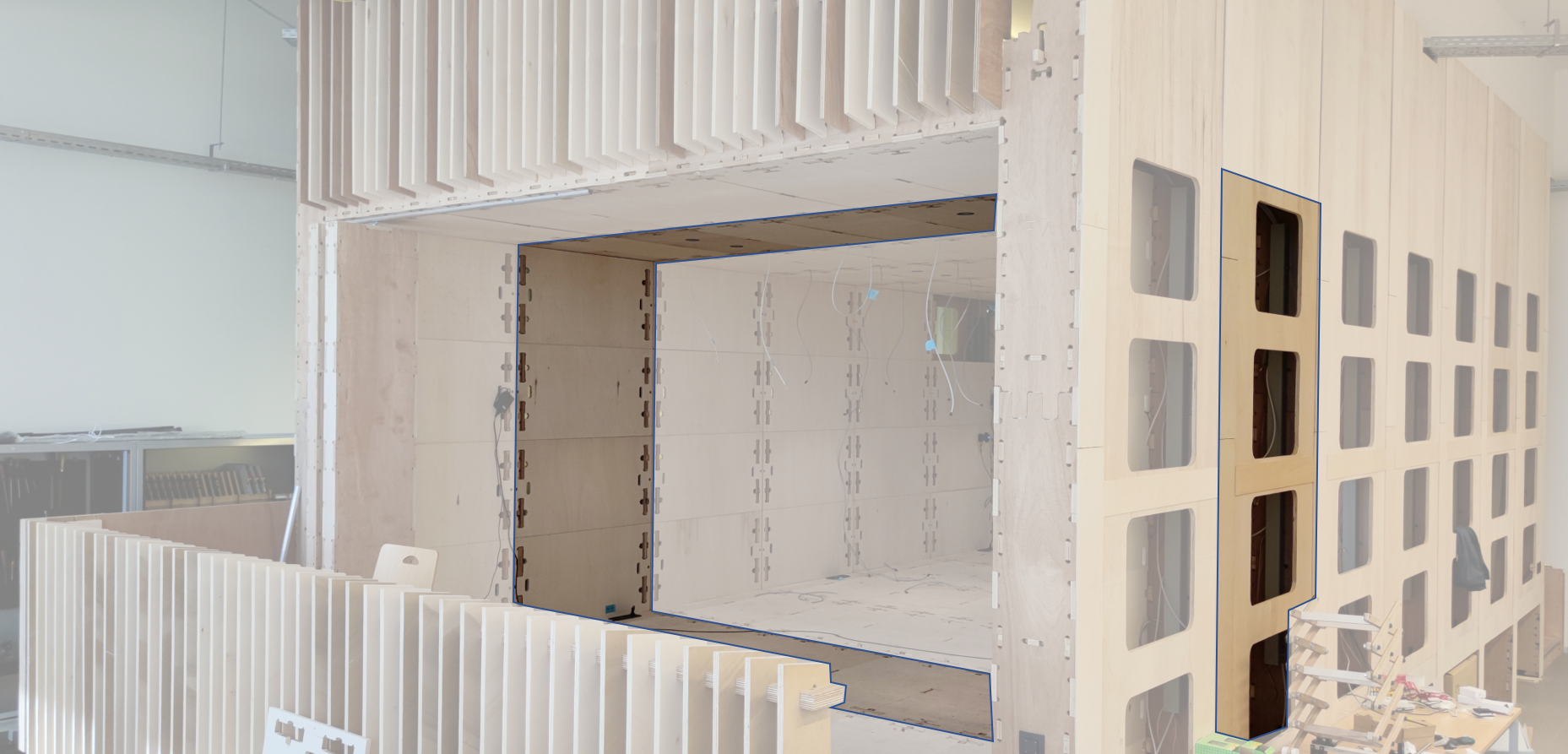}
    \end{subfigure}%
    \qquad
    \begin{subfigure}[t]{0.25\textwidth}
    \centering
    \includegraphics[height=2.2in]{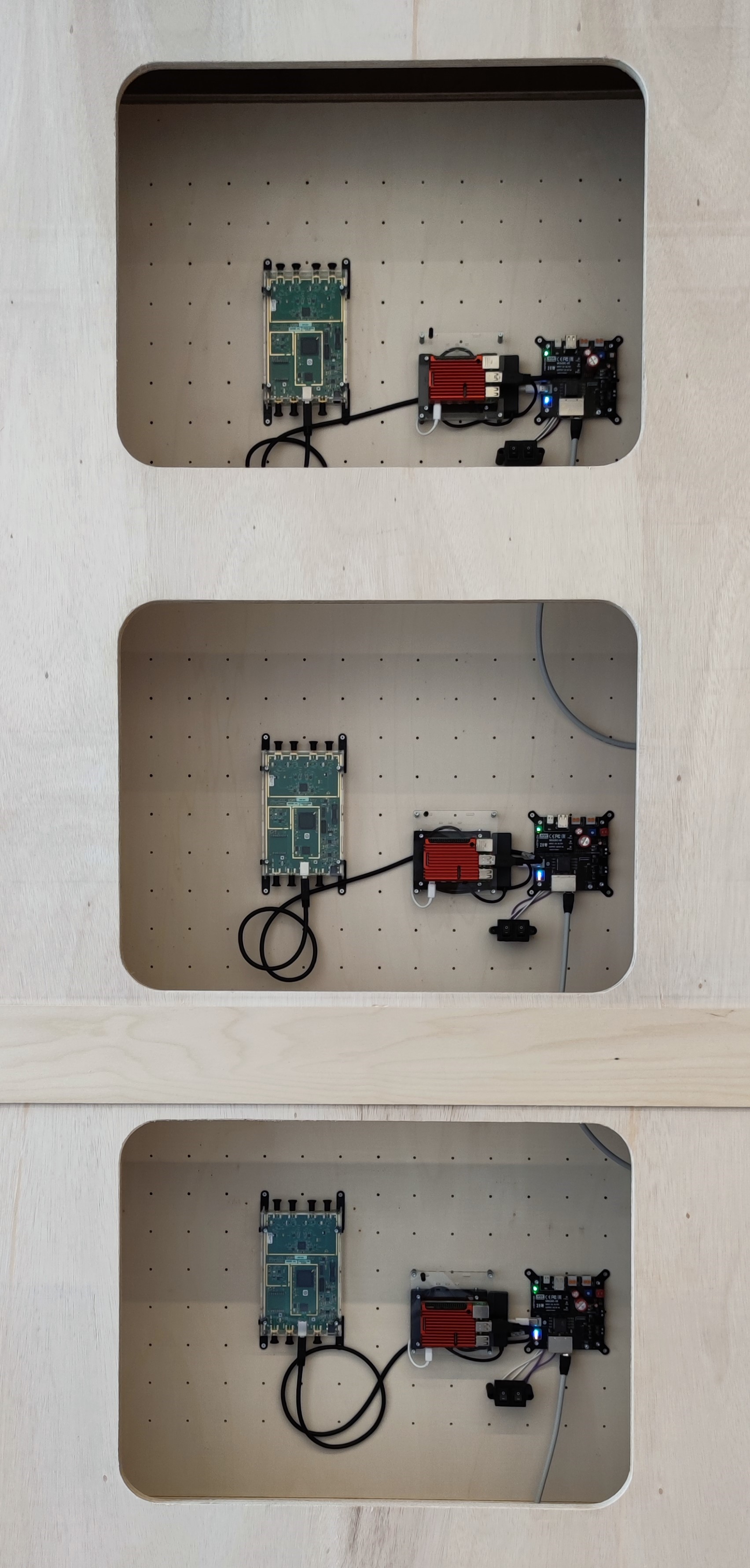}%
    \end{subfigure}
    \caption{\small Left: The Techtile support structure -- Right: The back of three tiles, equipped with the default setup, i.e., a software-defined radio (USRP B210), processing unit (Raspberry Pi 4) and power supply with Power-over-Ethernet. Each tile is connected to the central unit with an Ethernet cable, providing both power and data.}\label{fig:picture-techtile}
\end{figure}

The main technical challenges, in both the testbed and future distributed infrastructures, are the scalability and synchronization.  In contrast to other testbeds~\cite{7063446,8471108,10.1145/3117811.3119863}, this testbed does not utilize dedicated connections for communication and synchronization to each processing point, i.e., --in our case-- tile. The conventional approach to time synchronization requires all cables to have the same length, making deployment cumbersome and not scalable. In order to support high-speed connections between the tiles and the central processing, multiple hierarchies of processing must be organized to aggregate the high number of separate connections. To tackle these issues, all tiles are connected, powered (PoE++ IEEE802.3bt) and time synchronized (\acrshort{ptp} IEEE 1588) over Ethernet. By default, each tile has a processing unit, an \gls{sdr} and a power supply, as depicted in Figure~\ref{fig:picture-techtile}. This base configuration can be extended with custom solutions to support other use cases. 
To facilitate other research activities, all developed equipment and software is open-source available.\footnote{\url{github.com/techtile-by-dramco}}



In this paper we further provide more details on how the testbed is established in a modular way, and its features. The article is structured as follows. First, the support structure is discussed, focusing on the modular design of the detachable tiles. Thereafter, the backbone of the testbed is elaborated. It consists of a central processing unit or server and Ethernet connections to all tiles.  The on-tile equipment is introduced in Section~\ref{sec:ontile} with a focus on the default setup. In Section~\ref{sec:rover}, we discuss the rover supporting automated 3D sampling of the room, thereby reducing the time-consuming and labor-intensive task of conducting experiments manually. Lastly, the foreseen research and development activities are summarized in Section~\ref{sec:applications}. The list is by no means exhaustive and we welcome other R\&D activities and applications from industry and academic.

\FloatBarrier
\section{Techtile Construction -- Modular and Open}\label{sec:construction}
The Techtile support structure, illustrated in Figure~\ref{fig:picture-techtile}, is based on the WikiHouse\footnote{\url{www.wikihouse.cc}} concept. WikiHouse is an open-source building system. Such constructions are made of standardized wooden parts. Building further on the modular design, the walls, floor and ceiling of the structure are comprised of 140 detachable tiles. The two walls of the building support 28~tiles each. The ceiling and ground floor support 42 and 52 tiles, respectively. 
The room is \SI{4}{\meter} by \SI{8}{\meter} and is \SI{2.4}{\meter}.

A tile has a size of \SI{120}{\centi\meter} by \SI{60}{\centi\meter}. To mount the electronics, the tile features a \SI{5}{\centi\meter} by \SI{5}{\centi\meter} grid. The grid is made of M3 inserts, allowing conveniently installing components with standardized M3 screws. In addition to attaching equipment and custom electronics to the tile, we allow designing your own tile for specific applications. For instance, \gls{vlc} and \gls{vlp} systems using LED fixtures can permanently be mounted in the ceiling. We also support attaching equipment to the inside of the room by means of overlays. An overlay has the same grid structure as a tile and is mounted on top of the tile. This keeps the aesthetics, while still supporting attaching equipment inside. As the infrastructure is mainly designed for embedding electronics inside walls, we emphasize on only using the overlays if absolutely required. By using overlays, research demanding a \gls{los}, e.g., visible light and acoustics, are supported. Overlays are more flexible compared to designing your own tile, as overlays can be installed on all tiles.


\FloatBarrier
\section{Techtile Backbone -- Everything over Ethernet}\label{sec:backbone}
The backbone of the infrastructure consists of a central server which is connected to all tiles over Ethernet.
By means of \gls{poe} midspans, \SI{90}{\watt} of power is supplied to each tile. Furthermore, the Ethernet switches support the IEEE-1588 \gls{ptp}, enabling high-accuracy clock distribution to all connected devices. 
Hence, the testbed provides communication, synchronization and power over Ethernet.
As everything is connected over Ethernet, the developed system is easily scalable and is flexible in the manner in which devices are added and removed from the network. 


\subsection{Central Processing and Networking}
The central server (Dell PowerEdge R7525) has \SI{512}{\giga\byte} RAM, two NVIDIA Tesla T4 \SI{16}{\giga\byte} \glspl{gpu} and two AMD 7302 \SI{3}{\giga\hertz} \glspl{cpu}, running Ubuntu Server 20.04 LTS. 

Apache Kafka is used as a communication and processing platform between the server and the tiles. It follows the publish/subscribe paradigm. Events and streams are organized and stored in topics. Events can be thought of as files residing in a topic acting as a folder. Besides communication, Kafka support stream processing, database integrations, searching and querying data, management, logging and monitoring, making it very flexible to support future applications. Furthermore, data is generated and processed following a consumer/producer scheme, decoupling the systems. This allows to easily let the same data be processed by different clients and to dynamically plug-in and shutdown consumers. 

At the networking side, the Dell S4148 switch is used. It features 48 \SI{10}{\giga\bit} Ethernet ports with IEEE 1588v2 support. In Techtile, four switches are deployed with a total of 192 connections. A Dell Virtual Edge Platform (VEP), running VyOS, handles routing and firewall.





\subsection{Power-over-Ethernet}

In order to keep the cable management practical, power to the tiles is provisioned with \gls{poe} technology, where both data and power go over the same Ethernet cable. With the introduction of the latest standard, IEEE 802.3bt, a power of maximum \SI{90}{\watt} can be provided.
While this maximum is not required for the default setup, implementing the latest standard ensures a generic solution, supporting also high-power applications. 

Our \gls{poe} architecture consists of a~\gls{pse} and 140~\glspl{pd}, supporting all \gls{poe} standards, i.e., IEEE 802.3af (\acrshort{poe}), IEEE 802.3at (\acrshort{poe}+) and recent 802.3bt (\acrshort{poe}++). 
To provision the \gls{poe}, the PD-96XXGC series midspans from Microchip is selected as \gls{poe} injectors, supporting also the latest 802.3bt \gls{poe}++ standard. Each midspan has a 10/100/1000 Mbps data rate pass through. In total a \gls{poe} budget of approximately \SI{9}{kW} is available. 


\subsection{Synchronization-over-Ethernet}
In order to provide time synchronization for all the tiles, \gls{ptp} IEEE 1588v2 is used. This protocol achieves a clock accuracy in the sub-microsecond range --and depending on the network, configuration and version even sub-nanosecond. The protocol follows a master-slave architecture. The root time reference is hold by the grandmaster clock and distributed to the other clocks in the network. \acrlong{ptp} supports both L2 and UDP transport. It has an operation similar to \gls{ntp}, where the master and slave exchange messages to determine the path delays and correct their clocks accordingly. In \gls{ptp}, different profiles are defined, each having different configurations and requirements. Such profiles are tailored for specific application and are, for example, used in \gls{tsn}~\cite{8412458}. Different clocks are defined based on their capabilities. For instance, transparent clocks are network devices which alter the timestamps in the packets to remove the time spent in these devices, effectively making them transparent to the \gls{ptp} protocol. A boundary clock on the other hand, serves as both master and slave. It functions as a master at some connections, while synchronizing to a master on another port. One of the ongoing work is to tune the configuration of the protocol to achieve nanosecond accuracy and determine the impact on the processing capabilities of the \gls{rpi} and the load on the network.

\FloatBarrier
\subsection{Data Acquisition System}
The data acquisition system will provide panels with a synchronized analog data acquisition channel, able to sample at 1.25\,MS/s, a resolution of 16 bit, and for a total number of 192 channels.
This sample frequency and number of bits is sufficient to sample almost every sensor which allows us to move the state-of-the-art in sensor fusion and federation and underpins our aim for multi-modal sensing and positioning research. It includes sampling of full continuous audio streams (i.e., microphones recording audible or ultrasonic sound) and sensors registering high-bandwidth visible light communication. The high number of synchronized channels allows for setting up truly dispersed architectures, so large and distributed arrays can be obtained. 

The data acquisition system also has 48 synchronized 16-bit DAC output channels able to steer a variety of actuators. For example an array of speakers, ranging into the deep ultrasonic range, can be implemented in research on transmit beam forming and backscattering for fully passive mobile devices.
Additional possibilities are the generation of modulated signals and multiple access coding schemes to drive power LEDs in \gls{vlc} applications.

The combination of the data-acquisition system and the edge devices forms the infrastructure that is required to validate new concepts based on distributed nodes. By using the fully synchronized sensing capabilities of the central data acquisition and processing infrastructure, a baseline performance can be established. This baseline performance can act as a reference for more realistic scenario's where distributed edge devices collaborate in a loosely coupled (and varying) configuration.

\section{On-Tile -- Sensors, Radios, Processing, and Other Cool Stuff}\label{sec:ontile}
Each tile can accommodate a diversity of sensors and actuators, transmitters and receivers for radio - and other waves, host processing resources to add distributed intelligence in the environments, and potential other cool stuff that creative researcher may want to experiment with.
In this section we elaborate on the custom \gls{poe} board, the \gls{sdr} and the \gls{rpi}. The \gls{poe} board delivers power to the tile and the \gls{rpi} processes the \gls{sdr} signals and serves as a platform for edge computing.

\FloatBarrier
\subsection{PoE Board}

\begin{figure}[!htb]
      \centering
      \begin{subfigure}[t]{0.4\textwidth}
        \centering
        \def\svgwidth{0.8\linewidth}\fontsize{8pt}{10pt}\selectfont
\begingroup%
  \makeatletter%
  \providecommand\color[2][]{%
    \errmessage{(Inkscape) Color is used for the text in Inkscape, but the package 'color.sty' is not loaded}%
    \renewcommand\color[2][]{}%
  }%
  \providecommand\transparent[1]{%
    \errmessage{(Inkscape) Transparency is used (non-zero) for the text in Inkscape, but the package 'transparent.sty' is not loaded}%
    \renewcommand\transparent[1]{}%
  }%
  \providecommand\rotatebox[2]{#2}%
  \newcommand*\fsize{\dimexpr\f@size pt\relax}%
  \newcommand*\lineheight[1]{\fontsize{\fsize}{#1\fsize}\selectfont}%
  \ifx\svgwidth\undefined%
    \setlength{\unitlength}{566.92913386bp}%
    \ifx\svgscale\undefined%
      \relax%
    \else%
      \setlength{\unitlength}{\unitlength * \real{\svgscale}}%
    \fi%
  \else%
    \setlength{\unitlength}{\svgwidth}%
  \fi%
  \global\let\svgwidth\undefined%
  \global\let\svgscale\undefined%
  \makeatother%
  \begin{picture}(1,1)%
    \lineheight{1}%
    \setlength\tabcolsep{0pt}%
    \put(0,0){\includegraphics[width=\unitlength,page=1]{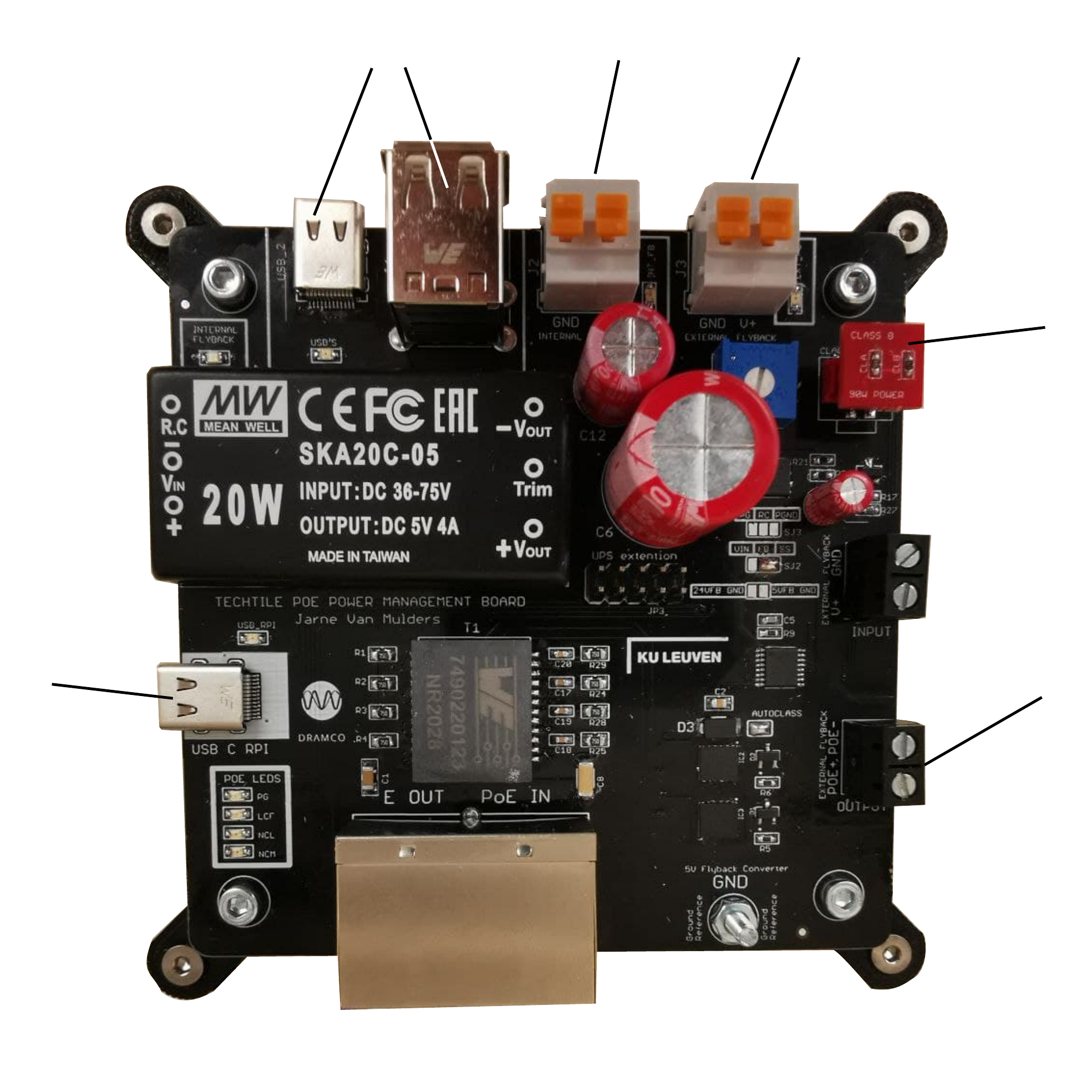}}%
    \put(0.33655361,0.95571518){\color[rgb]{0,0,0}\makebox(0,0)[lt]{\lineheight{1.25}\smash{\begin{tabular}[t]{l}A\end{tabular}}}}%
    \put(0.55316978,0.95669884){\color[rgb]{0,0,0}\makebox(0,0)[lt]{\lineheight{1.25}\smash{\begin{tabular}[t]{l}B\end{tabular}}}}%
    \put(0.72430328,0.95785637){\color[rgb]{0,0,0}\makebox(0,0)[lt]{\lineheight{1.25}\smash{\begin{tabular}[t]{l}C\end{tabular}}}}%
    \put(0.96087799,0.68627838){\color[rgb]{0,0,0}\makebox(0,0)[lt]{\lineheight{1.25}\smash{\begin{tabular}[t]{l}D\end{tabular}}}}%
    \put(0.96183762,0.35){\color[rgb]{0,0,0}\makebox(0,0)[lt]{\lineheight{1.25}\smash{\begin{tabular}[t]{l}E\end{tabular}}}}%
    \put(-0.00204408,0.35405296){\color[rgb]{0,0,0}\makebox(0,0)[lt]{\lineheight{1.25}\smash{\begin{tabular}[t]{l}F\end{tabular}}}}%
    \put(0,0){\includegraphics[width=\unitlength,page=2]{img/poe-board-labels.pdf}}%
  \end{picture}%
\endgroup%

        \caption{\small Custom \gls{pd} board. The different connectors are labeled and elaborated in (b).}\label{fig:poe-board-img}
    \end{subfigure}
    \hfill
    \begin{subfigure}[t]{0.55\textwidth}
         \centering
         \includegraphics[width=0.75\textwidth]{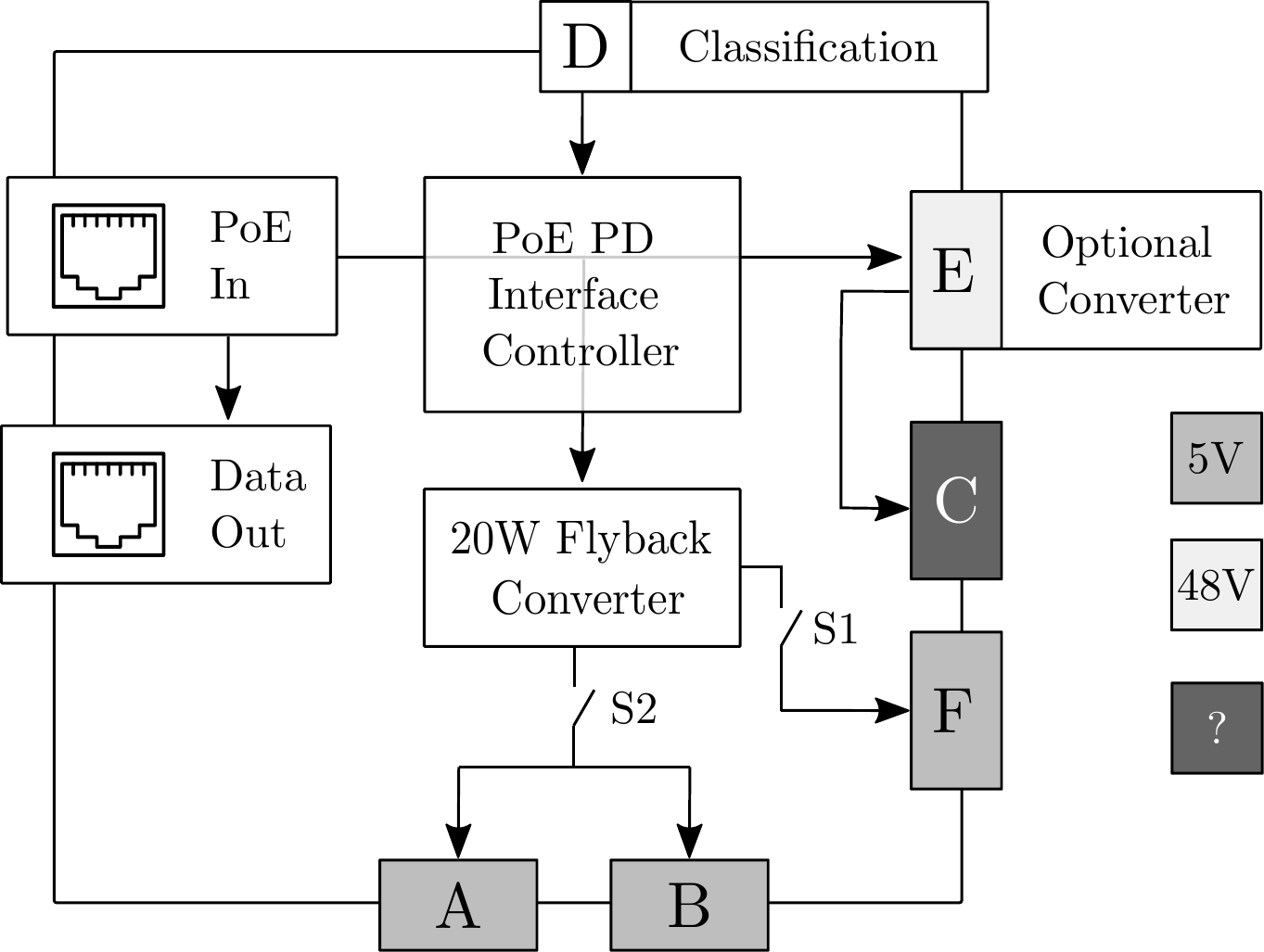}
         \caption{\small This block diagram represents the connection between the connectors, \gls{poe} controller and optional DC/DC converter. The output voltages available on the connectors are illustrated by different shades of gray. The optional DC/DC converter attached to E, can output different voltages on connector C. The switches S1 and S2 can be toggled at anytime.
         }
         \label{fig:poe-board-blockdiagram}
    \end{subfigure}
    \caption{\small Custom \acrlong{poe} \acrfull{pd}}
    \label{fig:poe-board}
\end{figure}

The standard \gls{poe} \gls{hat} for the \acrlong{rpi}, or other derivative forms,
are readily available but do not support the latest \gls{poe} version. For this reason, a 802.3bt supported \gls{pd} circuit was developed, as shown in Figure~\ref{fig:poe-board}. The system is based on the ON Semiconductor NCP1096 which is the PoE-PD interface controller. On top of the 802.3bt support, the board features several connectors and voltages to power different devices and sensors, as illsutrated in Figure~\ref{fig:poe-board-blockdiagram}. \Gls{poe} distributes \SI{48}{\volt} over Ethernet.

\textbf{Connectors and Usage.}
The \SI{48}{\volt} output is directly available on the board through terminal block C.  In case other voltages are required, an external DC/DC converter can be connected through terminal blocks E, and the ouput is available on terminal block C. The other output connections are connected to the internal flyback converter providing \SI{5}{\volt} with a maximum power of \SI{20}{\watt}. We adopted two bistable switches. One switch (S1) is dedicated to the USB-C connector for the \gls{rpi}, the other (S2) is used for the other USB-C connector, the two USB-A connectors and terminal block B. These switches prevent that the peripherals --and especially-- the \gls{rpi} will boot-up every time there is a power cycle. Furthermore, we can ensure that not all systems are powered, saving energy and mitigating any damage due to unintentional power outages. 

\textbf{Power Classification.}
The \gls{poe} \gls{pd} interface controller supports 8 different \gls{poe} classes, which are programmable with resistors. The classification is adjustable by means of replacing the classification board and consists of two resistors and a connector (Figure~\ref{fig:poe-board-img} [D]). The connector makes it convenient to switch the class rapidly although a power cycle is required. The most regular classes: 3 (\SI{13}{\watt}), 4 (\SI{25.5}{\watt}) and 8 (\SIrange[tophrase={--}]{71.3}{90}{\watt}) are provided. 
Besides a fixed power class, the class assignment is negotiated automatically. For instance, when only the \gls{rpi} and \gls{usrp} is connected, a power class of 3 is sufficient. However, as the class is determined at the beginning of power delivery --and fixed afterwards--, it is crucial that sufficient energy is extracted during start-up. Because the \gls{rpi} and \gls{usrp} power consumption and the tile setup is variable, it is not recommended to use autoclass, rather select the desired class in advance. Furthermore, as we have a finite power budget, knowing the consumption of each tile ensures that we adhere to the power constraint. Not respecting the agreed power, leads unfortunately to a power disconnection by the \gls{pse} and results in an abrupt power shutdown, causing potential damage and loss of measurement data.

\FloatBarrier
\subsection{Wireless RF Communication through Software-Defined Radio}
The testbed  hosts a fabric of distributed \glspl{sdr}. Each tile is equipped with one \gls{usrp} B210, featuring two RF channels. The B210 has a maximum transmit power of \SI{20}{dBm}.
This \gls{sdr} supports up to \SI{56}{\mega\hertz} of real-time bandwidth through the AD9361 direct-conversion transceiver. The B210 can operate over a frequency range of \SI{70}{\mega\hertz} to \SI{6}{\giga\hertz}, thereby covering most licensed and unlicensed bands. The B210 hosts an open and reprogrammable Spartan6 XC6SLX150 \gls{fpga}. The baseband signal is processed by the host, i.e., \gls{rpi}~4, using USB 3.0. GNURadio, supported by the B210, enables adopting and designing a high range of protocols and standards, e.g., IEEE 802.11 (Wi-Fi), Bluetooth, LoRaWAN, provided by the open-source community.
The B210 can be fed with an external \SI{10}{\mega\hertz} clock and \gls{pps} for synchronized operation. The reference clock is used to generate all data clocks, sample clocks and local oscillators. In addition, an external \gls{pps} signal can be used for time synchronization between the \glspl{sdr}. In this manner, it is possible to coherently transmit and receive on all \glspl{sdr}.

\subsubsection{Architecture}
\begin{figure}[!htb]
      \centering
        \centering
        \includegraphics[width=0.6\textwidth]{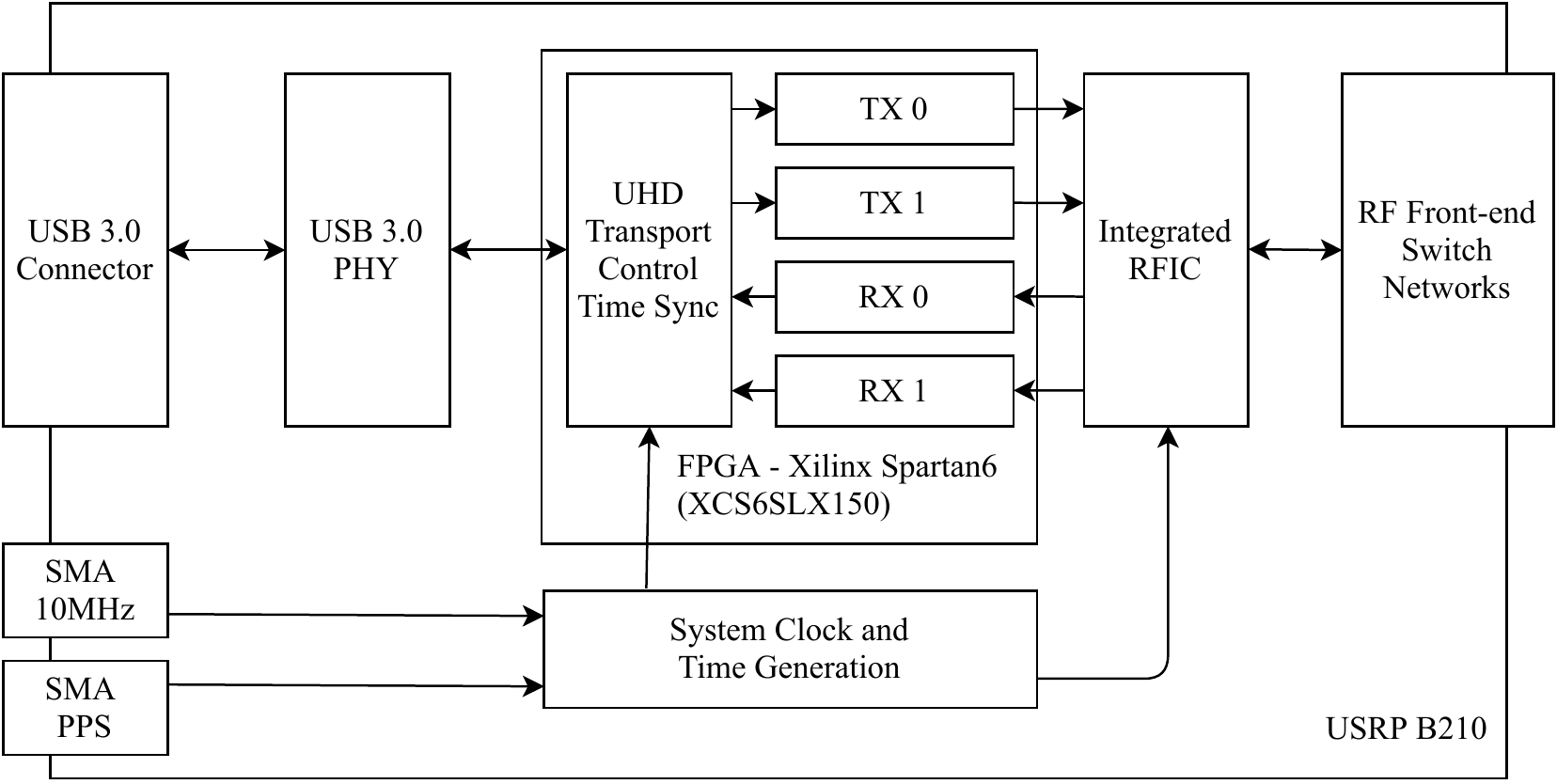}
    \caption{\small Simplified architecture the B210.}\label{fig:usrp}
\end{figure}

A simplified architecture of the USRP 210 is depicted in Figure~\ref{fig:usrp}.
It illustrates the connection with the host through the USB interface. The external time and frequency reference is used in the \gls{fpga} and \gls{rfic}.  
The RF frontend consists of mixers, filter, oscillators and amplifiers translating the signal from the RF domain to an \gls{if}. In receive mode, the complex baseband of that \gls{if} signal is sampled and clocked into the \gls{fpga}. National Intstruments has published the stock \gls{fpga} in open-source, and provides, o.a. digital down-conversion and filters for decimation. The resulting I/Q samples are transferred to the host though USB 3.0. The B210 features 2 RX/TX connections and 2 RX connections. The RX/TX connections are switched by an RF switch, thereby using only two identical transmit chains and two received chains sharing the same local oscillator for each TX or RX chain. 

\subsubsection{Processing the Baseband Signals}
The baseband signals are processed by the Raspberry Pi 4 and transfered over USB 3.0. Depending on the application, the distributed signals are aggregated to the central server for further processing. By having both local processing and central processing, different techniques regarding local, edge and cloud processing can be studied. 
Designing RF communication systems with the B210 is facilitated by the open-source \gls{uhd} from National Instruments. This library is supported in GNURadio and OpenBTS. 
In order to coherently transmit and receive, i.e., at the same time, over multiple USRPs, we output \SI{10}{\mega\hertz} and \gls{pps} with a custom \gls{pcb}. The \SI{10}{\mega\hertz} clock is derived from the \SI{54.0}{\mega\hertz} oscillator on the \gls{rpi} and tuned with the on-board \gls{pll}. By using a device tree overlay in Linux, the \SI{10}{\mega\hertz} signal is always present and independent on the state of the \gls{os}. The \gls{pps} is generated in a loadable kernel module.
The \gls{pcb}, installation procedure and drivers are available in open-source\footnote{\url{github.com/techtile-by-dramco/raspberrypi-sync}}.

\subsection{Edge Processing}
The default setup contains a \gls{rpi} capable of edge computing. 
To latest \gls{rpi} model, i.e., \gls{rpi}~4, is adopted. It has a rich feature set, such as \SI{8}{\giga\byte} of LPDDR4-3200 SDRAM and a powerfull processor, i.e., Quad core Cortex-A72 (ARM v8) 64-bit SoC 1.5GHz, tailored for high computing tasks.
For dedicated and more computation-intensive applications, custom edge computing platforms can be used, e.g., NVIDIA Jetson Nano, Google Coral and Intel NCS. 

To speed-up read/write operations, we adopted a \gls{ssd} opposed to the standard micro-SD card storage. Besides the speed improvement, the \gls{ssd} is less sensitive to power failure and has a higher capacity to price ratio. Furthermore, SD cards are designed to store data, while \glspl{ssd} are tailored for a high number of read/write operations, resulting from running an \gls{os}. These characteristics make \gls{ssd} a more robust and suitable technology.


\section{Automated 3D Sampling -- a Rover in the House}\label{sec:rover}

To speed-up measurements, reduce labor-intensive tasks and mitigate human-errors, we have designed a rover to perform automated 3D sampling inside the testbed. The rover consists of a baseplate, hosting the processing unit and batteries, and a scissor lift to move the measurement equipment to the required height.\footnote{This section contains work realized by two students, Arne Reyniers and Jonas De Schoenmacker, in the context of their master thesis.}

\textbf{Baseplate.}
The baseplate contains a Raspberry Pi running \gls{ros}. It controls the wheels, reads-out the sensors and performs navigation. By adopting \gls{ros}, the development of the rover is sped-up and is more easily extendable for future applications. \Gls{ros} consists of software packages dedicated for building robots, containing the necessary drivers and algorithms. The communication, employed in \gls{ros}, between nodes, i.e., processing entities, is handled by an asynchronous publish/subscribe message passing system. In this manner, \gls{ros} forces the developer to decouple different parts of the system. For synchronized communication, the request/response pattern is used.

A 16 cells battery pack, in a 4S4P configuration, powers the system. The battery voltage ranges from \SI{10}{\volt} to \SI{16.2}{\volt} depending on the charge state. The battery back, based on \gls{lco} technology, has a capacity of \SI{170}{\watt\hour} and can handle up to \SI{480}{\watt} peak power.  
Besides powering the rover, the battery pack provides power to the peripherals and measurement equipment in order to conduct the experiments. When the battery is low, the rover is able to navigate to its charging station, recharge and resume the measurements. 


\textbf{Navigation.}
Marvelmind indoor RTLS is used as a positioning system. It has a precision of \SI{2}{\centi\meter}~\cite{amsters2019evaluation}. Ultrasonic beacons are used to acquire the position of the mobile beacon (mounted on the baseplate). Four beacons are fixed in the Techtile infrastructure. The position is determined by trilateration. To filter out outliers, we make use of a Kalman filter. The recursive algorithm uses consecutive inputs to estimate the current state. Based on the uncertainties of the inputs and the uncertainties of the previous estimates, the current state is approximated. 
Besides localization, the rover features obstacle detection to mitigate crashing into objects in the room. Ultrasonic sensors mounted on the sides of the rover's baseplate are used and have a resolution of \SI{3}{\milli\meter} and a range of \SIrange{2}{400}{\centi\meter}. 

\begin{figure}[!htb]
     \hfill
      \begin{subfigure}[t]{0.48\textwidth}
         \centering
         \includegraphics[height=0.6\textwidth]{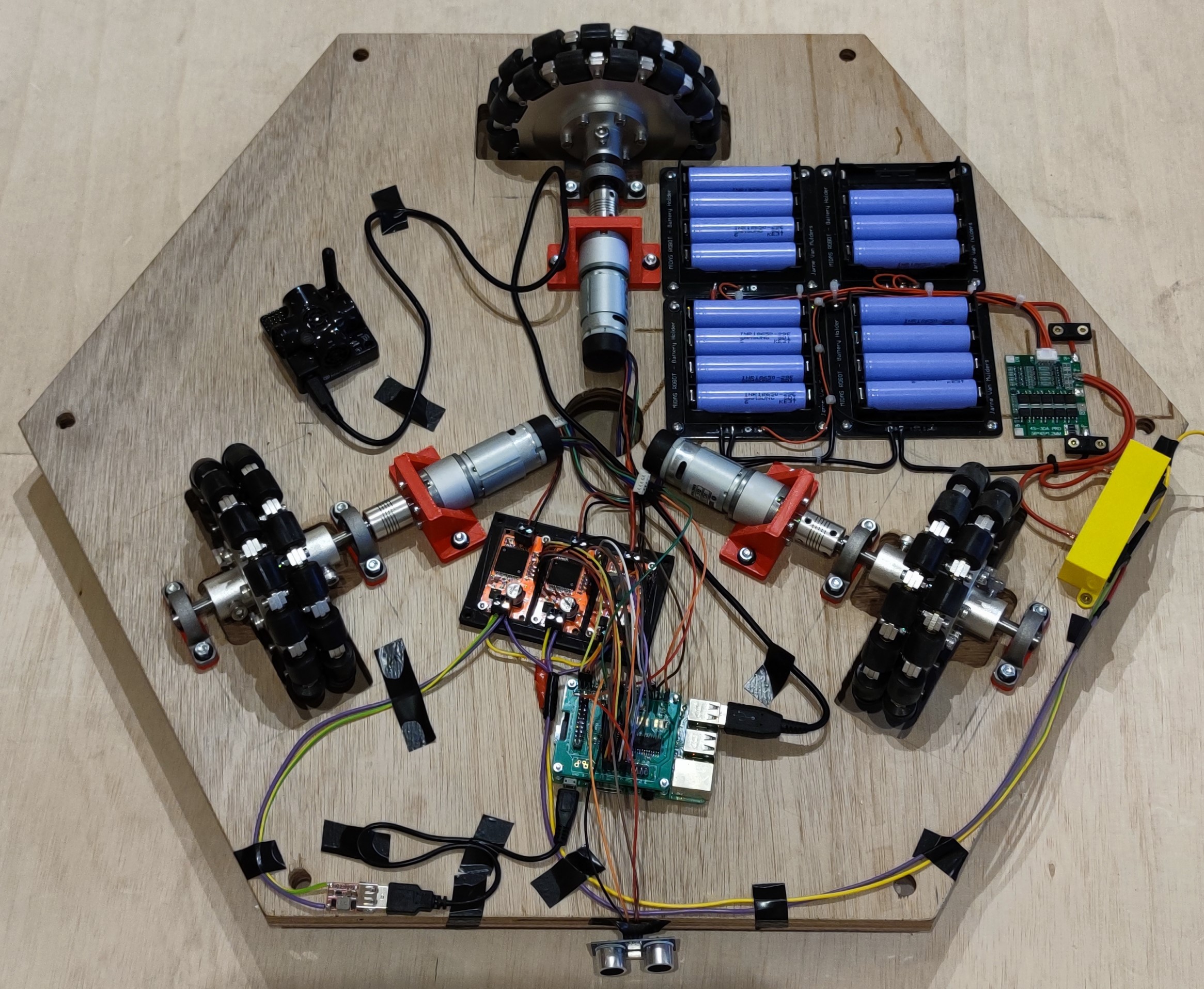}
         \caption{{\small Baseplate hosting the omnidirectional wheels, Controller (Raspberry Pi running \gls{ros}), battery pack and Marvelmind beacon.}}
    \end{subfigure}
    \hfill
    \begin{subfigure}[t]{0.48\textwidth}
         \centering
         \includegraphics[height=0.6\textwidth]{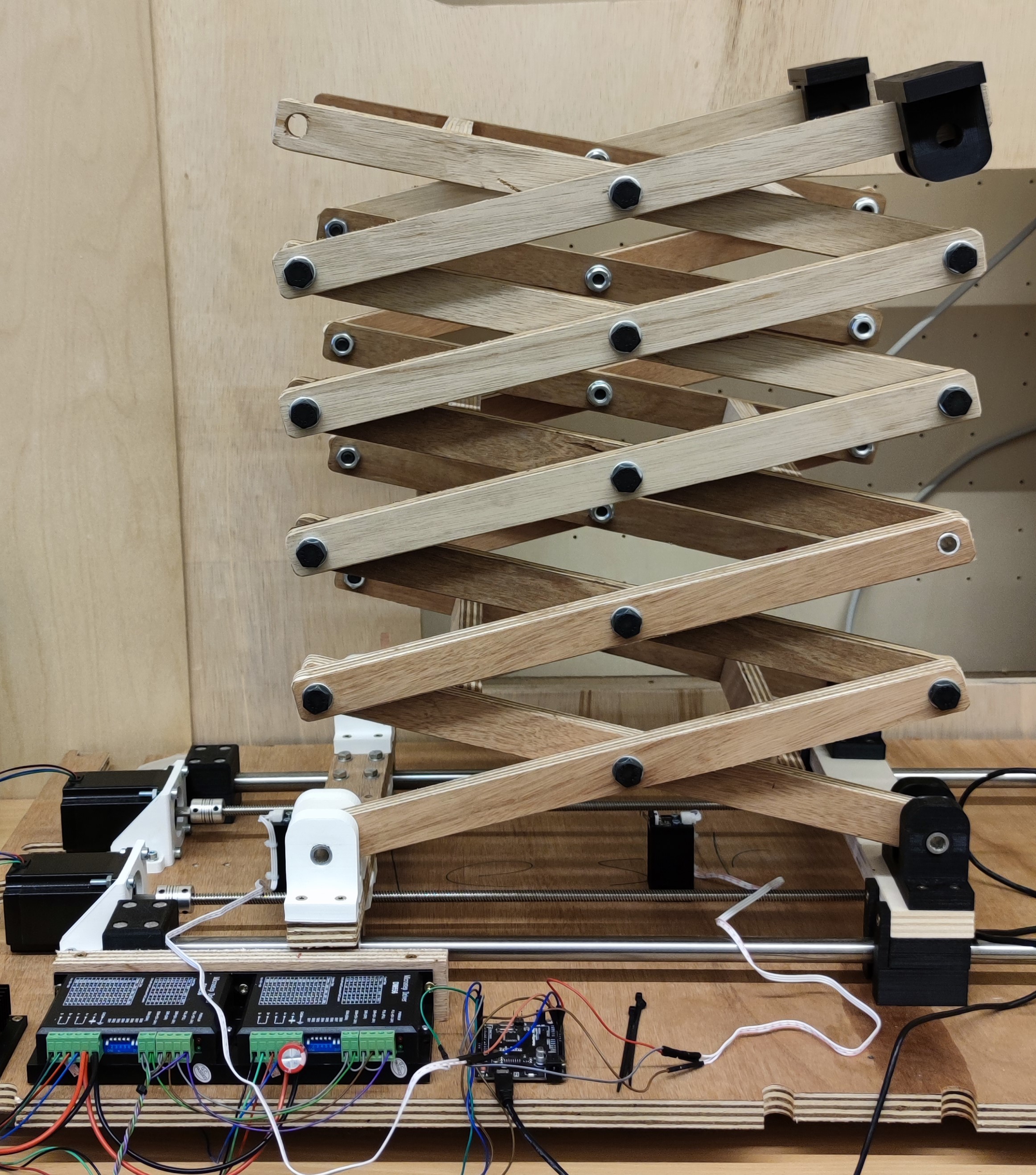}
         \caption{{\small Scissor lift having a range of 55-\SI{185}{\centi\meter}.}}\label{fig:lift}
    \end{subfigure}
    \hfill
    \caption{\small Rover to perform automated 3D sampling.}\label{fig:robot}
\end{figure}

\textbf{Scissor lift.}
The baseplate of the rover can navigate through the room in 2D space. By mounting the scissor lift (Figure~\ref{fig:lift}) on top of the baseplate, the rover is able to measure in 3D space. The lift has a range of \SI{55}{\centi\meter} to \SI{185}{\centi\meter}. The equipment can be mounted on top of the lift. Power is supplied to the equipment from the battery pack. The scissor lift can be controlled with AT-commands. The height of the lift is determined by the VL53L1X \gls{tof} sensor using a \SI{940}{\nano\meter} invisible Class~1 laser. An advertised accuracy and ranging error of $\pm$ \SI{20}{mm} in both ambient light and dark light conditions.

\FloatBarrier
\section{Research and Development Activities - Inviting Creative Experiments}\label{sec:applications}
The TechTile experimental facility has been designed to enable a wide range of experiments. We foresee, a.o., the following research and development activities: i) development of next-generation Internet-of-Things solutions, ii) experimental validation of beyond-5G communication, e.g., RadioWeaves, iii) positioning and sensing based on acoustic signals, iv) wireless charging and v) visible light communications and positioning.

\textbf{Internet-of-Things.}
We develop the next generation IoT solutions that should accommodate a massive number of low power devices and upgrade them to get smarter, leveraging AI approaches. We conceive auto-discovery, self-diagnostics, and recovery approaches, to ultimately achieve zero e-waste impact.

\textbf{(Beyond) 5G communication.}
We pursue experimental validation of (distributed) multi-antenna systems to support huge number of (low power) devices and provide ultra-reliable low latency connectivity. We will investigate cell-free operation and Large Intelligent Surfaces, new paradigms that may serve 6G systems.

\textbf{Acoustic sensing and indoor positioning.}
We sense acoustic signals for a better understanding of the environment. Through hybrid RF-acoustic signaling we pursue positioning of low energy devices with unprecedented accuracy.

\textbf{Secure connected devices.}
We exploit physical features of propagation and directivity to increase the security of connected devices, and enable private local networking solutions.

\textbf{Wireless charging.}
We investigate whether/how devices can get charged without the need for cables, and eventually ‘on their spot’.

\textbf{Visible Light Communication and positioning.}
We develop communication and positioning systems that modulate LED light sources without impacting their illumination functionality. They complement conventional RF systems where the interference poses aggravating problems.

\section*{Acknowledgments}
This project has received funding from the European Union's Horizon 2020 research and innovation programme under grant agreement No. 101013425.

The realization of the Techtile infrastructure is made possible thanks to the generous support of a bequest granted to the Science, Engineering and Technology Group of the KU Leuven. We have received hardware from both Niko and On Semiconductor. We also thank our equipment suppliers National Instruments, Dell and W\"urth Electronics for their special support for this pioneering development.

{\footnotesize \printbibliography}

\end{document}